# Steady State and Dynamics of Joule Heating in Magnetic Tunnel Junctions Observed via the Temperature Dependence of RKKY Coupling

A. Chavent[1, 2], C. Ducruet[2], C. Portemont[2], L. Vila[1], J. Alvarez-Hérault[2], R. Sousa[1], I.L. Prejbeanu[1] and B. Dieny[1], *Fellow, IEEE*

[1]Univ. Grenoble Alpes, INAC-SPINTEC, F-38000 Grenoble, France
CNRS, INAC-SPINTEC, F-38000 Grenoble, France
CEA, INAC-SPINTEC, F-38000 Grenoble
[2]Crocus Technology, F-38025 Grenoble, France

**Understanding quantitatively the heating dynamics in magnetic tunnel junctions (MTJ) submitted to current pulses is very important in the context of spin-transfer-torque magnetic random access memory development. Here we provide a method to probe the heating of MTJ using the RKKY coupling of a synthetic ferrimagnetic storage layer as a thermal sensor. The temperature increase versus applied bias voltage is measured thanks to the decrease of the spin-flop field with temperature. This method allows distinguishing spin transfer torque (STT) effects from the influence of temperature on the switching field. The heating dynamics is then studied in real-time by probing the conductance variation due to spin-flop rotation during heating. This approach provides a new method for measuring fast heating in spintronic devices, particularly magnetic random access memory (MRAM) using thermally assisted or spin transfer torque writing.**

In MRAM, a current is sent through a magnetic tunnel junction (MTJ) to switch the storage layer magnetization by spin transfer torque (STT) [1-3], or in the plane of the bottom electrode if the storage layer magnetization is switched by spin orbit torque (SOT) [4-6]. In all cases, the readout is performed by measuring at low bias voltage (~0.2V) the resistance of the MTJ which differs in parallel and antiparallel magnetic configuration due to the tunnel magnetoresistance phenomenon (TMR). Depending on the current amplitude, pulse duration and MTJ resistance, the current can also increase the temperature because of the power dissipated in the line or in the junction [7-9]. Solutions to reduce the current flow and the associated power consumption are currently under investigation, in particular by using the voltage control of magnetic anisotropy [10]. The switching behavior of spin torque driven devices [11-14] is greatly influenced by a temperature increase, possibly larger than 100°C even during pulses in the nanosecond range. The temperature variations affect the magnetic parameters, such as magnetization, magnetocrystalline anisotropy, and change the thermal activation energy. If the relationship between MTJ temperature and current intensity is unknown, an additional free parameter must be adjusted to fit the experimental switching phase diagrams and derive the spin torque efficiency. This unknown T(V) relationship cannot be addressed by simply comparing the coercivity dependence with temperature to that created by Joule heating because in most cases the current necessary for heating is large enough to induce spin-torque that impacts the coercivity of the device, as later detailed in this paper. Therefore, it is important to know the relation between temperature and voltage by other means. As an example, this has been recently investigated using the spin-wave thermal population as a temperature probe [15].

In Thermally Assisted MRAM (TA-MRAM) [7-9], the storage layer is pinned by exchange bias to an antiferromagnetic (AF) layer. When a current is sent through the MTJ, Joule effect heats the MTJ above the blocking temperature (Tb) of the AF. It is then possible to unpin the storage layer and set it in the desired direction thanks to either STT or a magnetic field.

Knowing the time variation of temperature under heating current pulse is of prime importance for TA-MRAM devices since it may be the limiting parameter for the speed. In fact, it depends on the switching regime. If the magnetic field or the STT is strong enough, the switching regime may be precessional. In that case, the switching time is limited by the gyromagnetic ratio to 0.2ns-1ns. For a lower energy supply, the regime is thermally activated. The characteristic switching time is then related to the energy barrier and temperature and may vary from 1ns to an arbitrary long time. Typically, for a practical TA-MRAM memory device, the switching time will have to be lower than 10ns. In that case, one must make sure that the heating duration is faster to avoid it from becoming the limiting factor in the device switching. This heating time has been calculated by thermal finite element simulations and was found to be in the range 2ns to 30ns for typical devices [16-17]. It strongly depends on the geometry of the pillar, and the materials buffering and capping the magnetic stack. For a power density of 50mW/µm², a time constant of 2.7ns was obtained in agreement with simulation. The time required for 90% of temperature increase is then 8ns (~3×2.7ns).

A simple 1D heating model gives a time dependence of temperature of the form $1 - e^{\frac{-t}{\tau}}$ where $\tau$ is the previously mentioned time constant. This exponential law does not describe the behavior of a real system with a complex geometry because the heat is dissipated through leads as well as the sides of the pillars. In that case, the time dependence is the superposition of different characteristic times [18].

In this work, we provide a solution to independently evaluate the heating in an MTJ whose free layer is driven by spin transfer torque, using the temperature dependence of

RKKY (for Ruderman Kittel Kasuya Yoshida) interlayer exchange coupling through Ru [19-20]. Later, the temperature dependence of RKKY coupling is used as a sensor in a real-time method to measure the temperature dynamics due to Joule heating.

The magnetic tunnel junction stack was realized by plasma vapor deposition (sputtering). It is composed of a PtMn pinned synthetic antiferromagnetic (SAF) reference layer comprising CoFe, Ru and CoFeB. The MgO barrier is a two-step naturally oxidized Mg, with RA=18±1 $\Omega.\mu m^2$. On top of the barrier, an exchange biased synthetic ferromagnetic (SFi) storage layer was depositied as shown in Fig.1. The MTJ TMR amplitude was measured by the current in-plane tunneling method and found to be 130%. Then, MTJ pillars were patterned by electron lithography followed by reactive ion etching of a Ta hard mask and ion beam etching of the MTJ itself. Electrical measurements were performed on circular and elliptical pillars of various sizes. The equivalent diameters are between 100nm and 200nm, and the aspect ratios are between 1 and 3.5.

Two different experimental methods were used, one for the static measurement of the temperature, and the other for the real time measurement of Joule heating dynamics. We will start by presenting the static measurements.

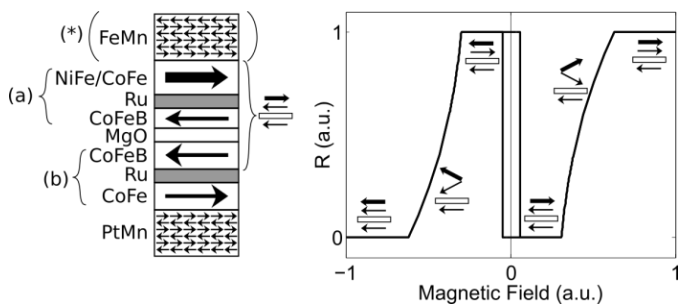

Figure 1: The magnetic stack is composed of (a) a SFi storage layer and (b) a SAF reference layer. The AF layer (*) is present in real-time measurements but not in the quasi-static measurements. A typical magnetic cycle of a free SFi layer is shown, whose thinner layer is adjacent to the barrier. Both spin-flop transitions cause resistance transition.

For the quasi-static temperature measurements, the free layer was: CoFeB / Ru / CoFe / NiFe, as shown in Fig. 1. The Ru provides antiparallel interlayer coupling at room temperature and the top composite layer (CoFe/NiFe) has a larger magnetic moment than the CoFeB layer, creating a synthetic ferrimagnet (SFi) storage layer.

The method consists in measuring the spin-flop field dependence versus bias voltage and also versus temperature using external heating. These two dependencies can then be used to determine the temperature dependence versus voltage. The spin-flop field is measured by R(H) loops at 5Hz, performed either at room temperature (20°C) and constant voltage (Fig. 2 (a)), or at constant temperature and low voltage (30mV), using a heating chuck (Fig. 2 (b)). For a magnetic field lower than 200Oe, we obtain the hysteretic cycle of the coherent SFi. For a magnetic field larger than 200Oe, the spin-flop is reached and causes the tunnel junction resistance to change because the layer in contact with the MgO barrier is the thinner of the SFi.

The spin-flop field is reached when the antiparallel configuration inside the SFi is no longer the state of minimum energy [21]. In practice, the magnetic moment of the thinner layer rotates by more than 90° at the spin-flop field, in our case the layer adjacent to the tunnel barrier. When this layer is oriented by 90° with respect to the reference layer, the resistance level is approximately C=40%, as shown on Fig. 1, where C is defined as:

$$C = \frac{R - R_{\min}}{R_{\max} - R_{\min}} \qquad (1)$$

The 40% level was used to determine the spin-flop field. The coercive field is defined here as the field at which the resistance level crosses C=50%. The spin-flop and coercive fields obtained that way from the room temperature R(H) loops are shown in Fig. 3.

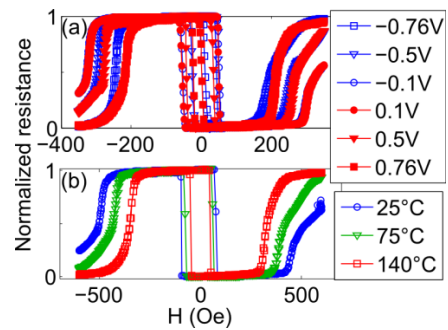

Figure 2: (a) R(H) cycles measured with DC voltage at room temperature. (b) R(H) cycles measured at a controlled temperature.

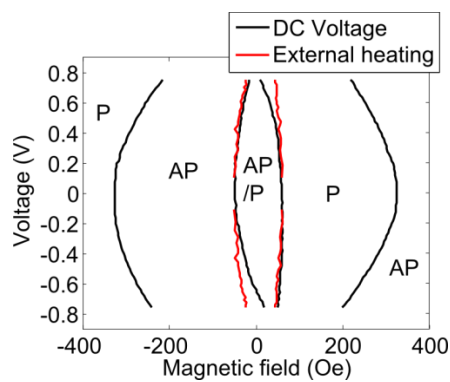

Figure 3: Diagram showing the boundaries of the coercive field and the spin-flop field. The red lines show the temperature effect of voltage on the coercive field, thanks to the temperature vs voltage relation calibrated with the spin-flop field.

An asymmetry in the coercive field between positive and negative polarity can be observed in Fig. 3. This is due to STT favoring the P (AP) state for positive (negative) polarity, corresponding to electron flow from the reference (storage) to the storage (reference) layer. Such an asymmetry exists also in the spin-flop field, but is not clearly visible in this diagram. All spin-flop fields (AP→P, P→AP, positive and negative voltage) are reported in Fig. 4 (a).

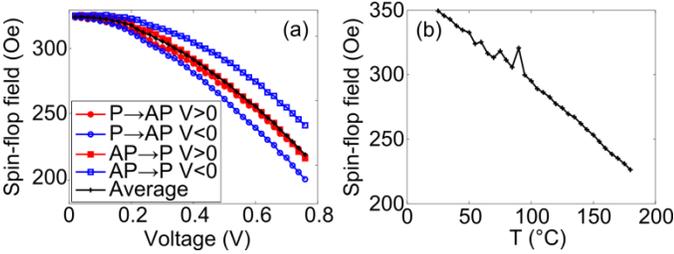

*Figure 4: (a) Spin-flop field at room temperature measured with R(H) loops at increasing DC voltage for each writing direction (P→AP and AP→P) and each voltage polarity. (b) Spin-flop field dependence with temperature, measured from R(H) loop obtained at 30mV bias voltage.*

We observe in Fig. 4 (a) that for P→AP, the spin flop field is lower for the negative polarity than for the positive polarity, and inversely for the AP→P reversal. This effect is attributed to STT. To eliminate the effect of STT in the spin-flop field reduction from the Joule heating contribution, the average of all four spin-flop field boundaries is calculated (positive and negative polarity and P→AP and AP→P switching), as shown by the black line in Fig. 4 (a). One should notice that the spin-flop fields of P→AP and AP→P switching are different. In fact, the AP state is favored which can be attributed to field-like STT [11,12]. The assumption that STT effects are cancelled out by averaging opposite polarities is justified, since the spin flop field characterizes a rotation process. As such, the equilibrium point will be affected by STT, as a small perturbation compared to the dominant antiferromagnetic exchange term. Therefore, a first order linear correction would effectively be cancelled out, since the STT term has opposite signs for opposite polarities.

The spin-flop field versus temperature measured at low voltage is plotted in Fig. 4 (b), showing a clear linear dependence with temperature. Extrapolation of the spin-flop field temperature dependence yields a vanishing spin-flop field at around 470°C, while in reality the approach to zero is expected to be asymptotic [19,20].

The temperature increase for a given voltage was calculated using the correspondence between spin-flop fields obtained as function of voltage (Fig. 4 (a)) and temperature (Fig. 4 (b)). The relation is obtained by normalization and interpolation. We observe in Fig. 5 (a) that the interpolated voltage dependence of temperature clearly follows a quadratic power law (without low temperature effect such as in [11]):

$$T - T_0 = \Delta T = \gamma V^2 \quad (2)$$

In this relation (2), the $\gamma$ coefficient depends on the stack electric resistance and heat conductivity. It is expressed in K/V². In Fig. 5 (b), the obtained heating coefficient is plotted versus the device surface $S$, showing clearly that heating is less efficient at smaller dot sizes. This is explained by the increasing heat loss through the pillar sidewalls $k_{\sqrt{S}}$, rather than by the top and bottom contacts $k_S$:

$$\gamma = \frac{1}{RA\left(k_S + k_{\sqrt{S}}/\sqrt{S}\right)} \quad (3)$$

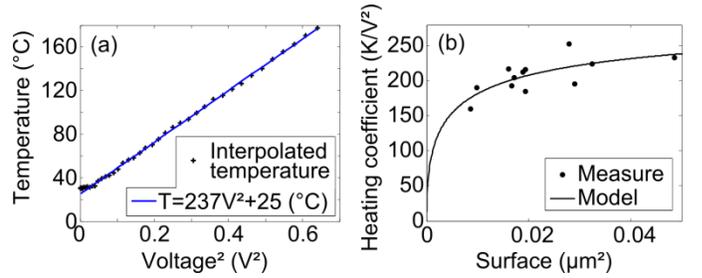

*Figure 5: (a) Temperature dependence versus voltage interpolated from spin-flop field dependence with temperature and voltage. (b) Heating coefficient $\gamma$ obtained from fit on different devices.*

Using this temperature-voltage dependence, it is now possible to include a temperature correction to the coercive field, to derive the STT effect contribution.

To illustrate this, we show in Fig. 3 the expected coercive field reduction versus voltage in red. The effect of STT becomes clear: Positive voltage reduces the switching field to P state compared to the thermal coercive field and negative voltage reduces the switching field to AP state, in agreement with the expected STT effect. When the unfavorable polarity is considered: positive voltage for AP state and negative voltage for P state, the switching field is much less affected by STT, and a possible increase of the switching field is not observed. This apparent contradiction can be reconciled assuming that the magnetization switching does not occur through a single reversal path, but rather through multiple possible paths. When STT favors the switching, lower energy barrier paths become possible reducing the coercive field. When STT provides additional stabilization, some reversal paths become less likely, but the barrier associated with others is not significantly altered resulting in an unchanged switching field.

In the second part of this paper, we report a study of the real time temperature dynamics in the MTJ submitted to heating current pulse. A pinned layer stack was used: CoFeB / Ru / CoFe / NiFe / FeMn, with the same SFi stack as before.

The idea is to set the system in a spin-flop state by applying an in-plane easy axis field and to observe the variation of angle of the magnetic moments due to the variation of RKKY coupling, using the change of resistance. (Fig. 6) By using a pinned storage layer, it is easier to keep the same initial state while sending multiple heating current pulses, as long as the blocking temperature is not reached.

The change of magnetoresistance was measured in real-time in a transmission set-up. A voltage pulse was applied using an Agilent 81134A pulse generator. The transmitted voltage was acquired with an 8GHZ bandwidth oscilloscope (Agilent Infiniium DSO 80804A) (Fig. 6). It is related to the resistance of the MTJ by:

$$V_{\text{transmitted}} = \frac{50\Omega}{R+100\Omega} 2V_{\text{set}}, \quad (4)$$

where $V_{\text{transmitted}}$ is the measured voltage, $V_{\text{set}}$ is the control voltage and $R$ is the resistance of the MTJ, around 800 $\Omega$.

The magnetic field was swept at 0.2Hz and a 50ns pulse was applied at a given magnetic field, while the voltage was acquired on the oscilloscope. This acquisition was made 50 times for each pulse voltage amplitude, and each magnetic field.

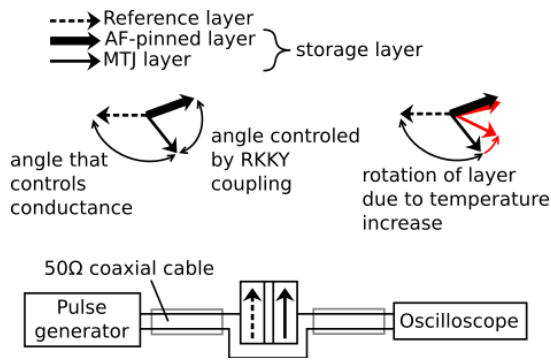

*Figure 6: Principle of the real-time measurement: the angle between reference layer and storage layer varies as RKKY coupling decreases with temperature increase. The angle variation is sensed trough the variation of resistance by transmission set-up.*

To start, the pulse amplitude was set to a moderate voltage of 550mV to keep the storage layer pinned. Unpinning events of the storage layer start appearing only at 600mV. At the same time, heating during the pulse reduces the RKKY coupling sufficiently to be observable in terms of resistance variation.

Oscilloscopes traces are acquired with different magnetic fields as shown in Fig. 7. The magnetic field direction was chosen, such as to switch the thinner layer of the SFi, through spin-flop at moderate fields, and to saturate it in parallel configuration with the second layer of the SFi at larger fields (440Oe – 460Oe). The 460Oe trace average was subtracted and the traces were normalized, as shown in Fig. 7. By normalizing, the time temperature dependence of TMR can be removed. The variation of TMR between 25°C and 180°C was about 30% at 30mV, while it was about 5% between 25°C and 150°C at 550mV.

With a low magnetic field of 100Oe, the MTJ stays in the same state P as without magnetic field. The low resistance P state corresponds to a high level voltage on the oscilloscope (Fig. 7). When the magnetic layer rotates, the voltage level decreases. In Fig. 7 (a), for a magnetic field amplitude of 240Oe, the starting voltage level is either the high level or an intermediate level. The two levels are possible because spin-flop is not a second order transition here, but a first order transition with a hysteretic cycle and has bistable region due to the short pulse duration and some anisotropy. With 240Oe, when the starting level is high, a transition generally occurs between 1ns and 15ns. This is a thermally activated switching. With 240Oe, when the initial state is the intermediate one, there is a gradual decrease of the voltage level, which becomes steady within a few nanoseconds.

In Fig. 7 (a) and (b), for the 280Oe field, the initial state is always an intermediate level which gradually decreases towards a steady level.

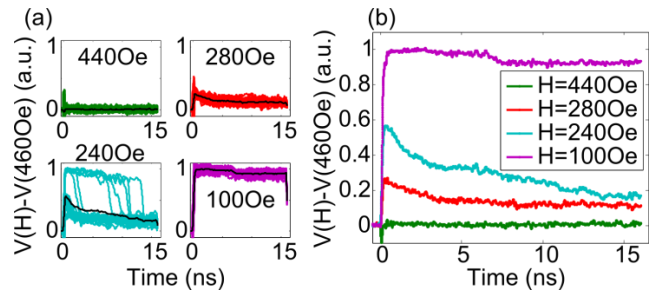

*Figure 7: (a) Normalized transmitted voltage traces, for varying fields around the spin-flop of an AF-pinned SFi storage layer. (b) Average. The applied voltage was 550mV.*

The change of behavior between 240Oe and 280Oe is attributed to the exit of the bistable region as shown in the sketch of Fig. 8 (b).

Characteristic individual transitions are shown in Fig. 8 (a) and located on the illustration of the assumed magnetic loop in Fig. 8 (b). At 240Oe, the junction is in a bistable state which causes both thermally activated switchings ① and a gradual reversal associated with gradual voltage transition due to thermal dynamics itself ②. When the magnetic field is increased up to 280Oe, the bistable region is left for a region where the state is not saturated. Further, at 440Oe, the magnetic state is saturated.

The gradual decrease of the average of individual traces obtained at 280Oe can be interpreted as the gradual decrease of the angle between the magnetization of the SFi as the RKKY coupling decreases with temperature. Since the reference layer magnetization stays fixed, it is only the magnetization rotation of the layer adjacent to the barrier that is observed.

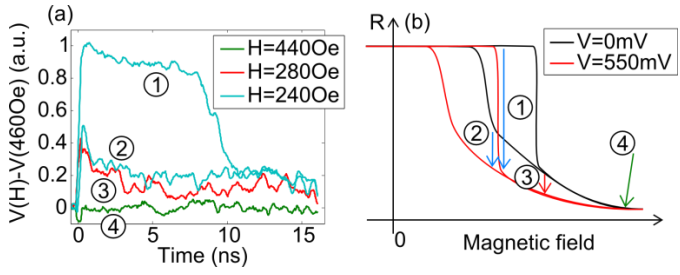

*Figure 8: (a) Examples of typical individual switching due to reduction of coupling energy with heating. (b) Sketch of the underlying magnetic loops at ambient temperature and temperature reached during the pulse, where the arrows show the transitions for each pulse example. At 240Oe, ① and ②: initial state is high or low due to bistability, then a transition to an intermediate state occurs. At 280Oe ③: all the traces are similar, beginning at an intermediate level and decreasing. At 440Oe, ④: saturation to low level.*

One has now to establish the link between the transmitted voltage level and the temperature. The intermediate physical phenomena which play a role in this relationship are the followings: the RKKY coupling $J_{RKKY}$ across the Ru spacer layer in the SFi free layer depends on temperature ([19,20], Fig. 4 (b)). Its variation changes the magnetization orientation of the magnetic layers, in particular the angle $\theta$ between the magnetic storage and reference layer magnetization [21]. The conductance $G$ of the MTJ is related to this angle [1] and affects the transmitted voltage $V_{transmitted}$ according to equation (4). There are four mathematical relations between these five parameters. Each parameter in these relations (magnetization, F/AF exchange, …) is known within some error bars. The solution lies in the fact that each relation is monotonous and continuous in the range of variation that we are considering ($\theta \in [\pi/2; \pi]$ $T \in [300K; 600K]$). It is then possible to work within a linear approximation since the variations of all the parameters are sufficiently small (about 15% of the full amplitude variation) [22]:

$$\delta V \propto \delta G \propto \delta\theta \propto \delta J_{RKKY} \propto \delta T \quad (5)$$

We normalized $V_{transmitted}$ for each step of time to remove the TMR variation with T. Within the assumption that the temperature follows an exponential law, we can fit the H=280Oe average trace as shown in Fig. 9, and extract a characteristic time of $\tau = 2.9$ns. This measure is in good agreement with the result obtained in a previous study [16] based on a pump-probe experiment. The critical time reported was $\tau = 2.7$ns [16]. This critical time is $\tau = C/k$, where $C$ is the heat capacity and $k$ is the thermal conductivity, so the agreement between experiments is expected because areal heat capacity and thermal conductivities are similar in the two sets of experiments.

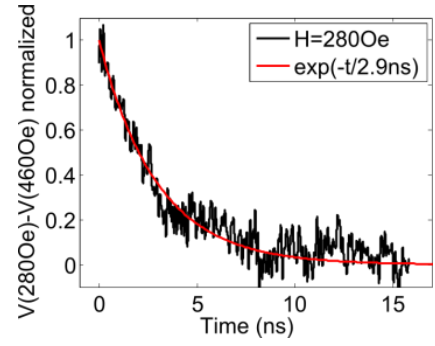

*Figure 9: In the case of H=280Oe, the transmitted voltage reproducibly relaxes from 0.3 to 0.1. The average of these traces is normalized here and fit with an exponential law related to heating.*

The approximation (5) has some limitations that should be discussed:
First, $J_{RKKY}(T)$ is mostly linear in the range of temperature we are considering, so for this first step of the measure chain, the linearization is a good approximation.
Then, one limit of this approximation is that for $\theta$ close to 0, $\pi/2$ or $\pi$, the linear term in the Taylor expansion is smaller than terms of higher order, and one should be careful and use the full expressions.
Finally, other parameters than RKKY coupling may depend on temperature and affect the observed real-time variation of conductance, especially $J_{ex}(T)$ the F/AF exchange energy that is well known to decrease in this range of temperature.

To conclude, this paper provides an estimation of the temperature increase using the temperature dependence of the antiferromagnetic exchange coupling through Ru in a synthetic ferrimagnetic storage layer. Switching of the free layer under STT can only be correctly modeled by including a temperature correction in the observed reversal field values, resulting in a reduction of the coercive field in agreement with the expected STT effect. When the effect of STT is to stabilize the existing state, no increase of the coercive field is observed. This might be expected assuming multiple reversal paths, whose energy barrier is lowered by STT or unaffected by it.

A real-time study of the reduction of RKKY coupling due to heating was also carried out. Under some conditions, the magnetic moment rotates rather than switches. This rotation can be related to a temperature variation, leading to a heating time constant of 2.9ns for an assumed exponential variation of the temperature.

ACKNOWLEDGMENT

This work received financial support from the ANR (French National Research Agency) under Project No. EXCALYB ANR-13-NANO-0010 and from ANRT (National Agency for Research and Technology), under CIFRE project No. 2012/1053.